\documentclass[sigconf]{acmart}

\AtBeginDocument{%
  }

\copyrightyear{2025}
\acmYear{2025}
\setcopyright{acmlicensed}
\acmConference[KDD '25] {Proceedings of the 31st ACM SIGKDD Conference on Knowledge Discovery and Data Mining V.2}{August 3--7, 2025}{Toronto, ON, Canada.}
\acmBooktitle{Proceedings of the 31st ACM SIGKDD Conference on Knowledge Discovery and Data Mining V.2 (KDD '25), August 3--7, 2025, Toronto, ON, Canada}
\acmISBN{979-8-4007-1454-2/25/08}
\acmDOI{10.1145/3711896.3737024.}

\usepackage{algorithm}      
\usepackage{algorithmic}    
\usepackage{amsmath}        

\usepackage{amssymb}        
\usepackage{subfigure}
\usepackage{enumitem}
\usepackage{graphicx}
\usepackage{booktabs}
\usepackage{colortbl}
\usepackage{xcolor}
\usepackage{multirow}
\usepackage{listings}
\usepackage{hyperref} 
\usepackage{placeins} 
\usepackage{booktabs} 
\usepackage{arydshln} 
\usepackage{xcolor}         

\usepackage{enumitem}  
\usepackage{longtable}  
\usepackage{multicol}  
\usepackage{fancyvrb} 
\usepackage[most]{tcolorbox}  
\newcommand{\coloredcircle}[2]{  
  \tikz[baseline=(char.base)]{  
    \node[shape=circle, fill=#1, inner sep=0.8pt, text=white, font=\bfseries] (char) {#2};}} 

\definecolor{color1}{HTML}{A6CAEC}
\definecolor{color2}{HTML}{B4E5A2}
\definecolor{color3}{HTML}{F6C6AD}

\newcommand{\ourmethod}{\textsc{LettinGo}~}

\definecolor{MySkyBlue}{RGB}{47,86,152} 
\usepackage{hyperref}       

\begin{document}

\title{LettinGo: Explore User Profile Generation \\for Recommendation System}

\author{Lu Wang}
\affiliation{%
  \institution{Microsoft Corporation}
  \city{Beijing}
  \country{China}}  
\email{wlu@microsoft.com}

\author{Di Zhang}
\authornote{Work done during internship at Microsoft.}
\affiliation{%
  \institution{Peking University}
    \city{Beijing}
  \country{China}}
\email{zhangdi@stu.pku.edu.cn}

\author{Fangkai Yang}
\affiliation{%
  \institution{Microsoft Corporation}
  \city{Beijing}
  \country{China}
}
\email{fangkaiyang@microsoft.com}

\author{Pu Zhao}
\affiliation{%
  \institution{Microsoft Corporation}
  \city{Beijing}
  \country{China}
}
\email{puzhao@microsoft.com}

\author{Jianfeng Liu}
\affiliation{%
  \institution{Microsoft Corporation}
  \city{Beijing}
  \country{China}
}
\email{jianfengliu@microsoft.com}

\author{Yuefeng Zhan }
\affiliation{%
  \institution{Microsoft Corporation}
  \city{Beijing}
  \country{China}
}
\email{yuefzh@microsoft.com}

\author{Hao Sun }
\affiliation{%
  \institution{Microsoft Corporation}
  \city{Beijing}
  \country{China}
}
\email{hasun@microsoft.com}

\author{Qingwei Lin}
\affiliation{%
  \institution{Microsoft Corporation}
  \city{Beijing}
  \country{China}
}
\email{qlin@microsoft.com}

\author{Weiwei Deng}
\affiliation{%
  \institution{Microsoft Corporation}
  \city{Beijing}
  \country{China}
}
\email{dedeng@microsoft.com}

\author{Dongmei Zhang}
\affiliation{%
  \institution{Microsoft Corporation}
  \city{Beijing}
  \country{China}
}
\email{dongmeiz@microsoft.com}

\author{Feng Sun}
\affiliation{%
  \institution{Microsoft Corporation}
  \city{Beijing}
  \country{China}
}
\email{sunfeng@microsoft.com}

\author{Qi Zhang}
\affiliation{%
  \institution{Microsoft Corporation}
  \city{Beijing}
  \country{China}
}
\email{qizhang@microsoft.com}

\renewcommand{\shortauthors}{Lu Wang et al.}

\begin{abstract}
User profiling is pivotal for recommendation systems, as it transforms raw user interaction data into concise and structured representations that drive personalized recommendations. While traditional embedding-based profiles lack interpretability and adaptability, recent advances with large language models (LLMs) enable text-based profiles that are semantically richer and more transparent. However, existing methods often adhere to fixed formats that limit their ability to capture the full diversity of user behaviors. In this paper, we introduce \textsc{LettinGo}, a novel framework for generating diverse and adaptive user profiles. By leveraging the expressive power of LLMs and incorporating direct feedback from downstream recommendation tasks, our approach avoids the rigid constraints imposed by supervised fine-tuning (SFT). Instead, we employ Direct Preference Optimization (DPO) to align the profile generator with task-specific performance, ensuring that the profiles remain adaptive and effective. \ourmethod operates in three stages: (1) exploring diverse user profiles via multiple LLMs, (2) evaluating profile quality based on their impact in recommendation systems, and (3) aligning the profile generation through pairwise preference data derived from task performance. Experimental results demonstrate that our framework significantly enhances recommendation accuracy, flexibility, and contextual awareness. This work enhances profile generation as a key innovation for next-generation recommendation systems.
\end{abstract}

\ccsdesc[500]{Information systems~Online advertising}
\ccsdesc[500]{Information systems~Recommender systems}

\keywords{Generative Recommender Model, User Preference Learning, Large Language Models}

\maketitle

\section{Introduction}

\begin{figure}[!t]
\centering
    \includegraphics[width=0.45\textwidth]{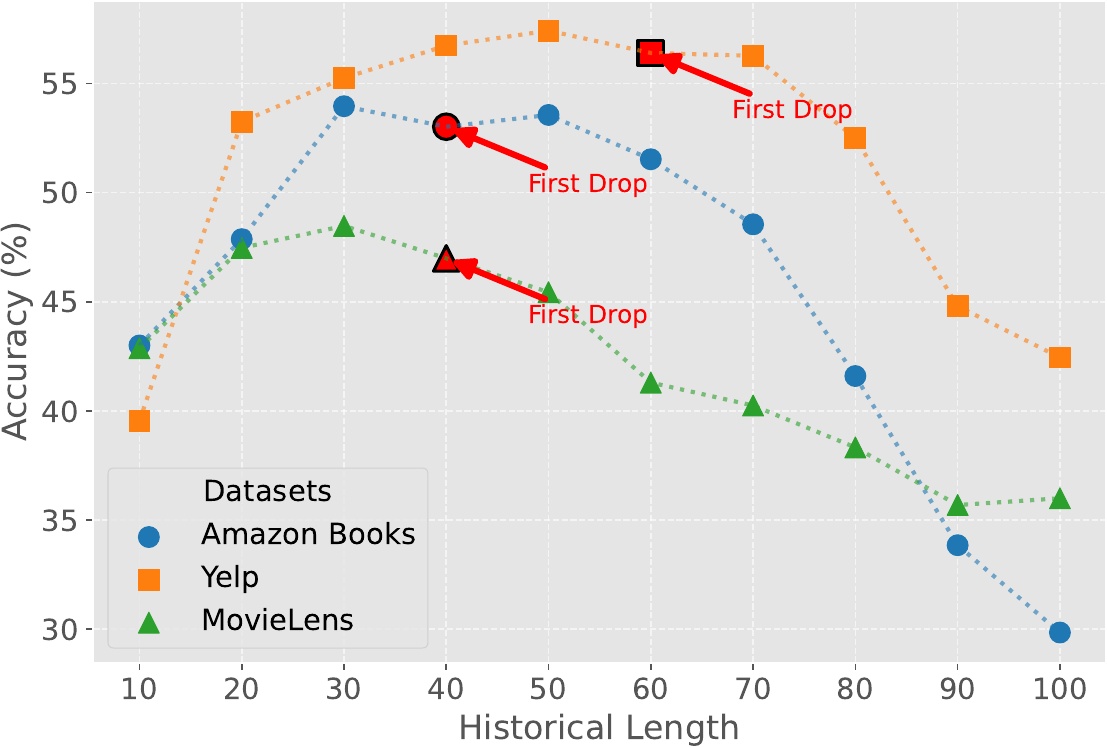}
\caption{Impact of Different Historical Textual Behavior Lengths on LLM (LLaMA 8B) Recommendation Performance for Amazon, Yelp, and MovieLens datasets. The performance improves initially with increased length but declines after a certain point. }
\label{fig:intro}
\vspace{-3mm}
\end{figure}

User profiling is a core element for modern recommendation systems, providing structured representations of user preferences and behaviors that drive personalized recommendations. By summarizing user interaction histories, profiles distill essential behavioral patterns from noisy raw data, enabling recommendation systems to deliver more accurate and relevant recommendations~\cite{wu2023computational}. This capability is critical in domains such as e-commerce, streaming services, and social media, where a nuanced understanding of user interests directly enhances satisfaction and engagement~\cite{wu2019session,purificato2024user}.

Traditional recommendation systems have primarily relied on embedding-based representations, encoding user preferences as numerical vectors in a latent space~\cite{chen2007content, widiyaningtyas2021user, vamosi2022deep}. While effective for similarity-based matching, these embeddings face notable challenges. They lack interpretability, making it difficult to understand the specific preferences being modeled, and they are difficult to update dynamically, especially in scenarios with sparse data or new users (commonly known as the cold-start problem)~\cite{gao2024rseiu, zhao2024rgmeta, rim2025cyclic, wang2025tensor, DBLP:journals/corr/abs-2411-00722}. Furthermore, these representations often fail to capture essential contextual signals, such as sequential patterns and temporal dependencies, which are critical for modeling complex user behavior~\cite{xia2022multi, huang2018csan, kulkarni2020context, chen2020learning}.

Recent advancements in Large Language Models (LLMs) have demonstrated strong potential in recommendation tasks, providing a new avenue for user profiling by leveraging natural language understanding to create semantically rich, interpretable, and adaptable profiles~\cite{zhang2024guided, kronhardt2024personaer, purificato2024user}. However, directly using a user's historical behavior as input to LLMs has proven insufficient for optimal performance. Token length constraints often lead to noise inclusion and the omission of critical information, limiting the effectiveness of recommendations. As shown in Figure~\ref{fig:intro}, the recommendation performance improves initially with increased input length but declines beyond a certain point due to these limitations.

These performance issues highlight the need for more structured and effective approaches to profile generation within LLMs. Existing methods typically prompt LLMs with rigid, predefined profile formats~\cite{yang2023palr, xi2024towards}, aiming to condense user information into manageable input sizes. While such profiles may improve efficiency in certain scenarios, they often fail to capture the complexity and variability of real-world user behavior. Consequently, static profiles struggle to generalize across diverse contexts and evolving task requirements. Defining what constitutes a ``good'' profile is inherently difficult, as its effectiveness is best measured by its impact on downstream recommendation performance. Profiles need to be dynamic and adaptable to better reflect the multi-dimensional nature of user preferences.

To address these issues, profiles should be adaptive, allowing for unconstrained representations that can incorporate real-time feedback from the recommendation task~\cite{molins2023embed, mendoza2024adaptive, liu2024beyond}. Existing supervised fine-tuning (SFT) methods impose further restrictions on profile formats~\cite{yang2024rag, wang2024muse}, which hinders flexibility. There is a growing need to explore profile generation approaches that are less constrained by predefined formats. Reinforcement learning (RL)-based strategies, such as DeepSeek-R1~\cite{guo2025deepseek}, offer a promising solution by enabling dynamic profile optimization and allowing the system to learn profiles directly from task performance without rigid format constraints.

In this paper, we introduce \ourmethod, a novel approach for exploring and generating adaptive user profiles to enhance downstream recommendation systems. It integrates exploration and feedback-driven alignment to produce profiles that more accurately reflect user behaviors and preferences. \ourmethod proceeds in three key stages:
\begin{enumerate}[nosep,leftmargin=*]
    \item \textbf{Profile Exploration:} We begin by collecting a diverse set of user profiles generated by various LLMs, including both closed-source models such as GPT-4o-mini~\cite{GPT-4o-mini} and open-source models that can be fine-tuned as profile generators. This stage aims to explore user profiles from a diverse and complementary set of profiles, while avoiding overfitting to specific profiles favored by LLMs to be fine-tuned as the profile generator.
    \item \textbf{Task-Driven Evaluation:} The generated profiles, combined with recent user interaction histories, are then integrated into a downstream recommendation system. This stage provides a direct evaluation of profile quality based on recommendation performance.
    \item \textbf{Profile Preference Alignment:} Finally, we leverage the evaluated profiles to construct pairwise preference data, which is used to fine-tune the LLMs. Compared with SFT methods that impose rigid constraints in profile formats, \ourmethod directly uses preference alignment to maintain flexibility in the generated profiles, letting go of the profile format and exploring good profiles for enhanced recommendation accuracy. 
\end{enumerate}


Our work focuses on advancing profile generation within LLM-based recommendation systems. Our key contribution lies in the development of an integrated framework that combines diverse profile exploration with task-driven optimization to learn a profile generator, which aims to generate adaptive and high-quality profiles. Experimental results demonstrate that \ourmethod significantly improves recommendation accuracy, adaptability, and contextual awareness. This work highlights the importance of enhancing user profile generation techniques for the next generation of recommendation systems.

\section{Related works}

\subsection{Profile in recommendation}
User profiles have long been a key element in improving the effectiveness of recommendation systems by tailoring suggestions to individual preferences and characteristics. Early approaches, such as CRESDUP~\cite{chen2007content}, utilize a client-side Dynamic User Profile (DUP) to deliver privacy-preserving personalized recommendations. Similarly, UPCSim~\cite{widiyaningtyas2021user} leverages user profile attributes—such as age, gender, occupation, and location—to compute correlation coefficients, thereby enhancing recommendation accuracy.

More recently, the emergence of LLMs has sparked growing interest in their application for personalized recommendations through the generation and refinement of user profiles. For instance, RLMRec~\cite{ren2024representation} employs GPT-3.5-turbo to create both user and item profiles, which are then aligned with collaborative filtering (CF) embeddings to mitigate feature noise. Likewise, KAR~\cite{xi2024towards} capitalizes on LLMs' reasoning abilities to produce user and item profiles, projecting them as additional features to boost performance in downstream recommendation tasks. The GPG method~\cite{zhang2024guided} proposes a prompt-based strategy, utilizing LLMs to generate profiles and incorporate contextual information for personalization. PALR~\cite{yang2023palr} constructs natural language user profiles based on interaction history, directly integrating these profiles into prompts for recommendation models. However, these works have not fully explored the potential of LLM-generated profiles or their impact on recommendation tasks.

Recent studies have revealed that while LLMs demonstrate promising capabilities in various domains, their direct application to recommendation tasks often yields suboptimal performance compared to specialized recommendation algorithms. This performance gap underscores the critical importance of domain-specific knowledge and collaborative filtering signals in recommendation systems~\cite{lin2023can}. To address this limitation, researchers have explored various approaches to integrate recommendation-specific collaborative signals into LLMs through parameter-efficient fine-tuning methods~\cite{liu2023pre}. Notable attempts in this direction include TALLRec~\cite{bao2023tallrec}, which employs the Low-Rank Adaptation (LoRA)~\cite{hu2021lora} architecture to fine-tune the LLaMA-7B model~\cite{touvron2023llama} on recommendation data. Similarly, Chen et al.~\cite{harte2023leveraging} investigated the fine-tuning of an OpenAI ada model for recommendation tasks. However, their findings indicate that directly fine-tuning LLMs for recommendation tasks still underperforms compared to alternative approaches, such as utilizing LLM embeddings for similarity matching or as initialization parameters for specialized recommendation models.

\subsection{LLM as Recommender itself}
The emergence of pre-trained large language models (LLMs) has brought significant success to the field of natural language processing (NLP), and LLMs have also shown immense potential in other domains, such as recommendation systems~\cite{wu2024survey}. While other works have already explored the use of natural language to explain the predictions of conventional recommendation models~\cite{xi2024towards,ren2024representation,li2023ctrl}, a promising direction is to leverage LLMs directly for recommendation tasks ~\cite{li2023large,yu2023self,dai2023uncovering,gao2023chat,hou2024large,wang2023zero}.Since fine-tuning LLMs requires substantial computational resources, zero-shot learning and in-context learning have gained widespread attention. Several studies have explored the use of LLMs as recommenders, yielding some initial successes ~\cite{gao2023chat,hou2024large,kang2023llms,liu2023chatgpt,wang2023zero}.For instance, ChatRec~\cite{gao2023chat} combines user profiles and interaction history to prompt the LLM for recommendations. \citet{kang2023llms} conducted a comprehensive comparison between LLMs and strong collaborative filtering (CF) methods. Their analysis revealed that, with fine-tuning, LLMs can achieve comparable or even superior performance with only a small portion of the training data, demonstrating the potential of LLMs in terms of data efficiency. \citet{wang2023zero} evaluate the zero-shot next-item recommendations performance of GPT-3, highlight the potential of using LLMs in zero-shot recommendation

\section{Method}

\subsection{Preliminary}
\noindent\textbf{Task Description.} The task is to leverage user's long interaction history to generate a profile, and then combined with user's short interaction history as the input to a recommendation system to decide if the user is interested in a given item. 

\noindent\textbf{Formulation.}
To be more specific, given a user $u$, their interaction history with $L$ interaction behaviors is defined as $\mathbf{H}_u=[h_{u,1}, h_{u,2}, \dots, h_{u,L}]$, where smaller subscription means more recent interaction behavior, \textit{i.e.}, $h_{u,l}$ is more recent than $h_{u,l+1}$. This interaction history is divided into two components:

\begin{itemize}
    \item \textbf{Recent History} \( \mathbf{h}_u = [h_{u,1}, h_{u,2}, \dots, h_{u,K}] \), which contains \( K \) most recent interactions utilized for the downstream prediction task.
    \item \textbf{Long History} \( \mathbf{H}_u = [h_{u,K+1}, h_{u,K+2}, \dots, h_{u,L}] \), which contains \( L-K \) historical interactions. This is used to explore and generate diverse user profiles in various formats.
\end{itemize}

The user profile \( \mathbf{p}_u \), representing a dynamic and flexible understanding of the user’s long-term preferences, is generated as:
\begin{equation}
    \mathbf{p}_u = f_{\text{LLM}}(\mathbf{H}_u),
\end{equation}
where \( f_{\text{LLM}} \) denotes the profile generation process conducted by an LLM. Unlike traditional fixed formats, the profile \( \mathbf{p}_u \) is unconstrained and can take various forms depending on the exploration step, enabling adaptability and expressiveness.

To evaluate the quality of the generated profiles, a downstream recommendation task is used. The task employs the generated profile \( \mathbf{p}_u \), the short history \( \mathbf{h}_u \), and a target item \( t \). These inputs are structured into a natural language prompt, forming the input to the downstream recommendation system:
\begin{equation}
    \hat{y}_u = f_{\text{Rec}}(\mathbf{p}_u, \mathbf{h}_u, t),
\end{equation}
where \( f_{\text{Rec}} \) represents the downstream recommendation system and \( \hat{y}_u \) is the predicted relevance score for the target item \( t \).

The feedback from the downstream task serves as an indirect measure of profile quality. By optimizing the profile generator based on this feedback, the model learns to align profile generation with downstream performance, ensuring higher-quality and context-aware profiles.

\begin{figure*}[t]
  \centering
  \includegraphics[width=0.95\linewidth]{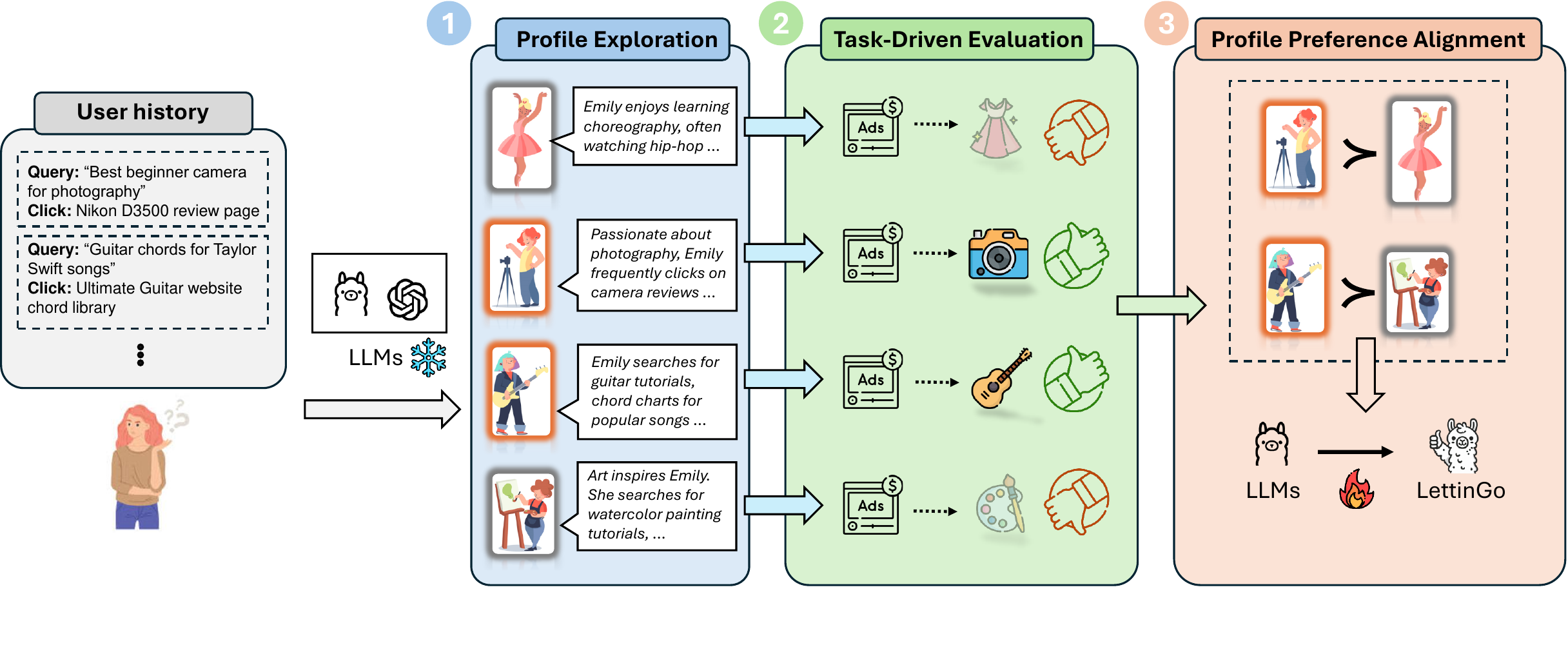}
  \caption{
  An overview of \textsc{LettinGo}, which consists of three stages. It begins with \textit{Profile Exploration}, where diverse user profiles are generated from existing LLMs. These profiles are then utilized in downstream recommendation systems for \textit{Task-Driven Evaluation}. Finally, the evaluated profiles are constructed as pairwise preference data to train the profile generator in \textit{Profile Preference Alignment}.
  }
  \label{fig:overview}
  \vspace{-3mm}
\end{figure*}

\subsection{The Design of \ourmethod}

We propose a novel method \ourmethod to explore and learn to generate good user profile for recommendation systems. It aims to create flexible, diverse, and adaptive representations of user behaviors and preferences, that help the downstream recommendation system to recommend appropriate items that suit the user's interests. As shown in Figure~\ref{fig:overview}, we first leverage various LLMs to explosively generate diverse user profiles in \coloredcircle{color1}{1}~\textbf{Profile Exploration} from user interaction and behavior history, capturing long-term behaviors and accommodating different contexts. Then, the generated user profile combined with the user's recent interaction history is fed into a recommendation system which evaluates the quality of the generated profile as \coloredcircle{color2}{2}~\textbf{Task-Driven Evaluation}. The evaluated user profiles form pairwise preference data that are used in \coloredcircle{color3}{3}~\textbf{Profile Preference Alignment} to train a good user profile generator. We discuss the details of each stage in the following sections.

\subsubsection{\textbf{Profile Exploration}} \label{sec:profile_generate}

Traditional fixed-format user profiles limits their adaptability to different tasks. To address this, we aim to explore diverse user profiles that reflect user preferences. By sampling profiles from various existing LLMs, we ensure a wide variety of user preference representations. 
To systematically explore diverse profile formats, We first design a simple, general-purpose prompt to instruct LLMs to generate user profiles:
    \begin{quote}
    \textit{
    \textcolor{MySkyBlue}{You will serve as an assistant to help me generate a user profile based on this user's sentiments history to better understand this users' interest and thus predict his/her sentiment about a target item. I will provide you with some behavior history of the user in this format:}
    \textcolor{teal}{[item attributes and sentiment]}
    \textcolor{MySkyBlue}{.The user profile you generate should contain as much useful content as possible to help predict the user's sentiment towards a new business.}\\
\textbf{\textcolor{brown}{USER HISTORY:}}
\textcolor{teal}{[user history]}.\\
\textbf{\textcolor{brown}{PROFILE YOU GENERATE:}}
} 
    \end{quote}

Profile generation is performed using both closed-source models such as GPT-4o-mini~\cite{GPT-4o-mini} and open-source models such as LLaMA~3~\cite{dubey2024llama}, where open-source models can be fine-tuned as profile generators. LLMs are typically pre-trained on large and diverse datasets, and they can offer robust, general-purpose representations of user preferences. Collecting profiles from diverse LLMs could avoid overfitting to specific profiles that preferred by the LLMs used as the base model to train as the profile generator.

Each LLM generates \( N = 10 \) profiles for a given user \( u \) based on their long history \( \mathbf{H}_u \). To encourage diversity in the profiles, we set a high sampling temperature (\( \text{temp} = 1.0 \)). The output is a collection of diverse profiles from all LLMs.

\subsubsection{\textbf{Task-Driven Profile Evaluation}}\label{sec:profile_evaluat}

Evaluating the quality of generated user profiles directly is challenging, as there is no predefined ground truth for a ``good'' profile. Instead, we use the performance of a downstream recommendation task as an indirect measure of profile quality. This allows us to assess profiles in a task-driven context, ensuring their relevance and effectiveness for personalized recommendations.

To evaluate profiles, we adopt a pairwise approach to construct labeled training data. For each user \( u \), the exploration phase generates multiple profiles \( \{\mathbf{p}_u^1, \mathbf{p}_u^2, \dots, \mathbf{p}_u^N\} \). We combine each profile with the user's recent interaction history \( \mathbf{h}_u \) and a target item \( t \), constructing a structured prompt for the downstream recommendation task:

\begin{quote}
\textit{
\textbf{\textcolor{MySkyBlue}{Given a user's past sentiments towards other items (sorted by time,from earliest to latest) in the format: }
\textcolor{teal}{[item attributes and sentiment]}
\textcolor{MySkyBlue}{, and a user profile which depict the user's interest about items, your task is helping me predict a user's possible sentiment about a target item based on these information in one word. The sentiment has three categories: like, neutral, and dislike. Remember, your output should only contain one word (like, neutral or dislike, in lowercase) that represent user sentiment you predict, without any additional content.}}\\
\textbf{\textcolor{brown}{USER HISTORY:}}
\textcolor{teal}{[user history]}.\\
\textbf{\textcolor{brown}{USER PROFILE:}}
\textcolor{teal}{[user profile]}.\\
\textbf{\textcolor{brown}{The candidate item is:}}
\textcolor{teal}{[item]}}.\\
\end{quote}

The downstream recommendation model \( f_{\text{Rec}} \) evaluates each profile\footnote{We omit $i$ in $\mathbf{p}_u^i$ which represents the $i^{th}$ profile of user $u$ for simple notation.}  \( \mathbf{p}_u \) and produces a relevance score \( \hat{y}_u \) for the target item. Profiles are then ranked based on the recommendation system's success in accurately predicting relevant items. We construct each pairwise training data by selecting two profiles for each user: A \textbf{positive profile} \( \mathbf{p}_u^{+}\), which contributes to a successful recommendation, and a \textbf{negative profile} \( \mathbf{p}_u^{-}\), which leads to a failed recommendation.  
The resulting pairwise dataset \( \{(\mathbf{p}_u^+, \mathbf{p}_u^-)\} \) provides a task-driven signal for profile quality, laying the foundation for the preference alignment stage.

\subsubsection{\textbf{Profile Preference Alignment}}

A primary goal of our method is to explore high-quality user profiles without fixed formats. We avoid limiting the format of user profiles to capture the diversity and complexity of user behaviors, ensuring adaptability to various downstream tasks and recommendation needs. This flexibility allows LLMs to fully utilize their expressive capabilities, generating nuanced and context-aware profiles that improve system performance and generalizability. Standard SFT methods inherently constrain the profile generator to specific formats seen during training, reducing flexibility. In addition, It is inherently difficult to define a ``ground truth'' profile for a user, as the optimal profile depends on how well it performs in the downstream recommendation task. This dependency necessitates a feedback-driven approach, aligning profile generation with the specific needs of the downstream model.

To ensure that the generated profiles remain flexible and task-specific, we train the profile generator using a feedback-driven alignment approach. Instead of relying on SFT approaches, which impose rigid constraints on profile format, we align the generator with task performance using the pairwise data from the task-driven evaluation stage. This approach optimizes the generator to produce high-quality profiles while maintaining the flexibility of the format.

We train the profile generator \( f_{\text{LLM}} \) using Direct Preference Optimization (DPO) with triple data \( (\mathbf{H}_u, \mathbf{p}_u^+, \mathbf{p}_u^-) \), where \( \mathbf{H}_u \) is the user's long interaction history. The generator's objective is to assign higher preference scores to positive profiles than to negative ones, aligning the profile generation process with downstream task performance.

The DPO~\cite{rafailov2023direct} loss is defined as:
\begin{equation}
    \mathcal{L}_{\text{DPO}} = -\mathbb{E}_{(\mathbf{H}_u, \mathbf{p}_u^+, \mathbf{p}_u^-)} \left[ \log \sigma \left( f_{\text{LLM}}(\mathbf{p}_u^+ \mid \mathbf{H}_u) - f_{\text{LLM}}(\mathbf{p}_u^- \mid \mathbf{H}_u) \right) \right],
\end{equation}
where \( \sigma(x) = 1 / (1 + e^{-x}) \) is the sigmoid function. This loss encourages the generator to produce profiles that maximize downstream recommendation performance.

The output of this stage is a fine-tuned profile generator \( f_{\text{LLM}} \) that dynamically generates user profiles optimized for downstream recommendation tasks, balancing flexibility and recommendation effectiveness.

\subsection{Data Collection Algorithm}
Algorithm~\ref{alg:dpo-multi-opt} presents our optimized data collection pipeline for preference-based fine-tuning. The goal is to construct a training dataset that allows the model to learn to distinguish high-quality user profiles from suboptimal ones, thereby aligning profile generation with downstream task objectives.

For each user $u$ in the dataset, we first generate multiple candidate profiles using a diverse set of LLM-based profilers $\{f_{\text{LLM}}^{(m)}\}_{m=1}^M$, each sampled $N$ times with a specified temperature $T$ to ensure variability. Each generated profile $\mathbf{p}$ is then evaluated using a downstream predictor $f_{\text{Rec}}$, which predicts user labels $\hat{\mathbf{y}}$ based on the profile and interaction history $\mathbb{H}_u$. Profiles that yield correct predictions ($\hat{\mathbf{y}} = \mathbf{y}_u$) are labeled as positive examples $\mathcal{P}_u^+$, while incorrect ones are considered negative $\mathcal{P}_u^-$.

If both positive and negative profiles are available for a user, we construct pairwise preference data by forming all possible $(\mathbf{p}_u^+, \mathbf{p}_u^-)$ pairs. These comparison tuples, consisting of the user’s history, the preferred and dispreferred profiles, and the ground-truth label, are added to the training dataset $\mathcal{D}_{\text{DPO}}$.

This procedure ensures that the collected training data captures fine-grained distinctions between good and bad profiles, as judged by their downstream utility. The resulting dataset is well-suited for Direct Preference Optimization (DPO), enabling the profile generator to improve by learning from task-relevant, model-informed feedback rather than human-annotated preferences.

\begin{algorithm}[H]
\caption{Optimized Data Collection}
\label{alg:dpo-multi-opt}
\begin{algorithmic}[1]
\REQUIRE User set $\mathcal{U}$; \\
Profilers $\{f_{\text{LLM}}^{(m)}\}_{m=1}^M$; \\
Downstream predictor $f_{\text{Rec}}$; \\
User interaction history $\mathbb{H}_u$; \\
Ground-truth labels $\mathbf{y}_u$ for each user $u$; \\
Sampling temperature $T$; \\
Profiles per model $N$.
\ENSURE Training dataset $\mathcal{D}$.

\STATE $\mathcal{D}_{\text{DPO}} \gets \emptyset$
\FOR{each user $u \in \mathcal{U}$}
    \STATE $\mathcal{P}_{u}^{+}, \mathcal{P}_{u}^{-} \gets \emptyset, \emptyset$ \COMMENT{Good/bad profiles for $u$}
    \FOR{each model $m = 1, \dots, M$}
        \FOR{$n = 1, \dots, N$}
            \STATE $\mathbf{p} \gets f_{\text{LLM}}^{(m)}(\mathbb{H}_u; T)$ \COMMENT{Generate profile}
            \STATE $\hat{\mathbf{y}} \gets f_{\text{Rec}}(\mathbf{p}, \mathbb{H}_u)$ \COMMENT{Predict}
            \IF{$\hat{\mathbf{y}} = \mathbf{y}_u$}
                \STATE $\mathcal{P}_{u}^{+} \gets \mathcal{P}_{u}^{+} \cup \{\mathbf{p}\}$
            \ELSE
                \STATE $\mathcal{P}_{u}^{-} \gets \mathcal{P}_{u}^{-} \cup \{\mathbf{p}\}$
            \ENDIF
        \ENDFOR
    \ENDFOR

    \IF{$\mathcal{P}_{u}^{+} \neq \emptyset$ \textbf{and} $\mathcal{P}_{u}^{-} \neq \emptyset$}
        \FOR{each $\mathbf{p}_{u}^{+} \in \mathcal{P}_{u}^{+}$}
            \FOR{each $\mathbf{p}_{u}^{-} \in \mathcal{P}_{u}^{-}$}
                \STATE $\mathcal{D}_{\text{DPO}} \gets \mathcal{D}_{\text{DPO}} \cup \{(\mathbb{H}_u, \mathbf{p}_{u}^{+}, \mathbf{p}_{u}^{-}, \mathbf{y}_u)\}$
            \ENDFOR
        \ENDFOR
    \ENDIF
\ENDFOR
\STATE \textbf{return} $\mathcal{D}$
\end{algorithmic}
\end{algorithm}

\section{Evaluation}
In this section, we present the experimental evaluation results of our framework across multiple datasets, addressing the following research questions:
\begin{itemize}
    \item RQ1: How does \ourmethod perform compared to traditional fixed-format profiles in recommendation systems? 
    
    \item RQ2: How does the feedback-driven alignment strategy through DPO contribute to profile generation quality?
    \item RQ3: What is the impact of historical interaction length on profile generation quality and downstream recommendation performance? 
    \item RQ4: How does \ourmethod enhance recommendation interpretability and capture diverse user preferences? 
\end{itemize}
\subsection{Experiments Settings}
\subsubsection{Datasets}

We conducted experiments on three widely-used recommendation datasets: Movielens-10M~\cite{Movielens-10M}, Amazon Books~\cite{ni2019justifying,Amazon-Books}, and Yelp~\cite{Yelp}, all of which consist of user ratings and reviews for items. To ensure meaningful analysis with sufficient interaction history, we filtered the datasets to retain only users with more than 70 historical interactions. 
For the test set, following the methodology of \citeauthor{kang2023llms}, we selected the last interacted item from users with more than 70 historical interactions and randomly sampled 2,000 users to construct the test set. When evaluating on the training set, the length of historical records used to generate profiles can be explicitly controlled (e.g., 30, 50, or 70 interactions). 
Due to computational constraints, we further sampled 3,000 users from the remaining data to construct the training set for profile generation. To ensure diversity in the training data and enable the profile generation model to effectively handle varying lengths of historical interactions, we divided the sampled users into three groups of 1,000 each. For each group, we used interaction histories of lengths 30, 50, and 70, respectively, as input to sample profiles. This ensured that the final training data included profiles generated from interaction histories of various lengths.

For each user, we generate 10 profiles, which were labeled and paired according to the methodology outlined in Section~\ref{sec:profile_evaluat}. Specifically, the profiles were paired based on whether they resulted in accurate or inaccurate downstream recommendation. The statistics of the training and test sets are presented in Table~\ref{tab:dataset}.

\begin{table}[t]
\caption{Statistics of the experimental datasets. }
\resizebox{0.5\textwidth}{!}{
\begin{tabular}{l|r|r|r|r|l}
\toprule
\textbf{Datasets} & \textbf{\#Users }& \textbf{\#Items} & \textbf{\#train} &\textbf{\#test }
& Features \\
\hline
Movielens-10M& 71567& 10681& 15,637  &2,000  
& Title, Genre \\
Amazon-Books & 1,850,187 & 483,579 & 13,212  &2,000  
& Title, Category\\
Yelp& 19,800& 22,734& 14,427  &2,000  &  Business name, Category\\
\bottomrule
\end{tabular}
}
\label{tab:dataset}
\vspace{-3mm}
\end{table}

\subsubsection{Evaluation}
Similar to GPG~\cite{zhang2024guided} and the approach in~\cite{kang2023llms}, we directly leverage LLMs and textual data to perform zero-shot user preference prediction. Following the experimental paradigm of Multi-Behaviour Recommendation~\cite{chen2023survey}, we evaluate the effectiveness of generated user profiles by framing it as a classification task. For each user, we input the following components into the downstream LLM recommendation system:User’s recent history,
A generated profile, and A candidate item.
The system predicts the label for the candidate item, classifying it into one of three categories: dislike, neutral, or like. The objective is to assess how well the input profile supports accurate predictions. If the system correctly classifies the item, the profile is validated as effective. If the prediction is incorrect, the profile is considered less representative of the user's preferences. We employ two widely-adopted classification metrics: Accuracy, weighted-F1 score. Given that our task involves three-way classification (like, neutral, and dislike), these metrics provide a comprehensive evaluation of the model's performance across all sentiment categories. The use of macro-averaging ensures that each class contributes equally to the overall evaluation, regardless of class imbalance.

\subsubsection{Baselines}
To enable a comprehensive comparison and to demonstrate the effectiveness of our proposed approach, we implement the following baseline methods. These include the most basic method, which predicts based solely on the most recent 10 interaction records~(\textbf{10H}), as well as other approaches that leverage natural language prompts to assist the model in generating user profiles using 30 long interaction history~(\textbf{30P}) for recommendation tasks:

\begin{itemize}

    \item \textbf{Prediction with 10 Recent History~(10H).} Following the experimental setup of prior research \cite{kang2023llms}, we adopt the use of the most recent 10 interaction records for direct prediction as a baseline. It is widely acknowledged that recent interaction records are the most indicative of a user's short-term preferences, making the most recent 10 interactions particularly significant for capturing immediate user intent.
    
    \item \textbf{KAR}~\cite{xi2024towards}. As mentioned in our introduction, increasing the number of interaction records can improve recommendation performance to some extent. However, excessively long interaction histories may introduce noise and lead to overly lengthy contexts, which can hinder model effectiveness. Compared to our method of summarizing user profiles, this approach directly utilizes the same number of interaction records for prediction.
    
    \item Both \textbf{RLMRec}~\cite{ren2024representation} and \textbf{PALR}~\cite{yang2023palr} are designed to prompt LLMs to generate user profiles for recommendation systems by combining recent interactions with long-term behavioral data. While the underlying approach of both methods is similar, they differ in the specific prompts used to guide the LLM in profile generation. To fairly compare the impact of profile generation on recommendation performance, we implement both baselines using only their respective prompts while keeping other components constant. The profile generation prompts of the baselines are in Appendix~\ref{appendix:baselineP}.
\end{itemize}

\subsubsection{Implementation Details}
In our experiments, we primarily employ LLaMA~\cite{dubey2024llama,touvron2023llama} as both the recommendation model and the profile generation model. Specifically, LLaMA3 8B Instruct is utilized as the downstream prediction model, while LLaMA3 8B Instruct~\cite{dubey2024llama} and LLaMA2 14B Chat~\cite{touvron2023llama} are used for generating user profiles. Additionally, we leverage GPT-4o-mini and Claude~\cite{claude} to generate diverse profiles. For the item recommendation, the model temperature is set to 0 to ensure deterministic outputs. In contrast, during profile sampling for training data collection, the temperature is increased to 1.0 to enhance diversity. The training process is implemented using the LLaMA-Factory framework~\cite{zheng2024llamafactory}. Key hyperparameters, such as batch size and learning rate, are determined through grid search to achieve optimal performance.

\subsection{Performance Comparison}

\begin{table*}[h!]
\caption{Performance on three dataset using LLaMA 3 (8B) and LLaMA 2 (13B).}
\centering
\begin{tabular}{lcc|cc|cc|cc|cc|cc}
\toprule
\textbf{Method} & \multicolumn{6}{c}{\textbf{LLaMA 3 (8B)}}& \multicolumn{6}{c}{\textbf{LLaMA 2 (13B)}}\\
\cmidrule(lr){2-7} \cmidrule(lr){8-13}
 & \multicolumn{2}{c}{Movie-lens}& \multicolumn{2}{c}{Yelp}& \multicolumn{2}{c}{Amazon Books}& \multicolumn{2}{c}{Movie-lens}& \multicolumn{2}{c}{Yelp}& \multicolumn{2}{c}{Amazon Books}\\
\cmidrule(r){2-13} 
                 & \textbf{Acc} & \textbf{F1} & \textbf{Acc} & \textbf{F1} & \textbf{Acc} & \textbf{F1} & \textbf{Acc} & \textbf{F1} & \textbf{Acc} & \textbf{F1} & \textbf{Acc} & \textbf{F1} \\
\midrule
10H                & 44.95& 41.08& 37.45& 42.92& 48.15& 55.93& 44.95& 41.08& 37.45& 42.92& 48.15& 55.93\\
KAR\cite{xi2024towards}& 51.00& 49.91& 67.90& 61.60& 67.40& 69.84& 48.00& 46.51& 65.70& 60.93& 60.05& 65.28\\
RLMRec\cite{ren2024representation}& 48.00& 47.07& 61.30& 58.96& 63.60& 67.42& 45.65& 45.26& 62.60& 58.90& 59.85& 65.05\\
PALR\cite{yang2023palr}& 48.10& 47.26& 64.10& 59.47& 66.40& 69.29& 49.10& 47.94& 63.10& 58.85& 60.10& 65.47\\
\hdashline[1pt/1pt]
\textbf{Ours}& & & & & & & & & & & & \\
10H+30P& 51.80& 51.05& \underline{70.70}& 62.91              & 70.55& \underline{71.79}& 50.45& 49.25& \underline{66.35}& \underline{60.47}& 64.75& 68.34\\
 10H+50P& 52.80& 51.16
& 70.65& \underline{63.17}& \underline{70.75}& 71.71& 51.05& 49.59& 65.65& 60.31& \underline{65.75}&\underline{69.01}\\
10H+70P& \underline{53.00}& \underline{51.69}& 70.40 & 62.94& 70.40 & 71.27 & \underline{51.09}& \underline{57.69}& 66.00& 60.14& 63.95& 67.71\\
\bottomrule
\end{tabular}
\label{table:main_result}
\end{table*}

\subsubsection{Improvement over baselines~(RQ1)}Table~\ref{table:main_result} presents the comparative results of our proposed methods against the baseline across three widely-used datasets: Movielens-10M, Amazon Books, and Yelp. We utilize LLama3-8B-Instruct as both the profile generator and the downstream recommendation model; during prediction, we keep the temperature at 0. 
To validate the effectiveness of our framework, we compare it against three state-of-the-art methods that enhance recommendation performance through profile generation. The experimental results are reported in Table~\ref{table:main_result}. Specifically, we use the prompts from these methods to generate profiles and include a baseline that predicts recommendations solely based on the most recent 10 interaction records. The experimental results demonstrate that our proposed method achieves significant performance improvements across all datasets. On the LLaMA3 8B Instruct model, our approach improves the accuracy by an average of 20 percentage points across the three datasets compared to the baseline that uses only historical interactions. Furthermore, compared to other profile-based approaches, our method exhibits a clear advantage, particularly on the Amazon dataset, where it achieves an accuracy of 66.30\% and an F1 score of 69.04\%, outperforming all baseline methods. This performance superiority can be attributed to the following key factors:
(i) Structured and informative profiles: By leveraging carefully designed prompt templates, \ourmethod generates more structured and information-rich user profiles, effectively capturing the core preference characteristics of users.
(ii) Dynamic and balanced preference modeling: By limiting the historical interactions to the most recent 10 records while incorporating the generated profile information, our method balances short-term and long-term user interests. This allows it to retain users' latest dynamic preferences while leveraging the stable, long-term preferences distilled in the profiles.
(iii) Task-aligned profile optimization: Through downstream task-driven evaluation, we select and align profile generation with the data most beneficial to downstream tasks. This alignment training enables the profile generator to consistently produce more effective and task-relevant profiles.

The experimental results also reveal differences in performance improvements across datasets, which may be attributed to the characteristics of each dataset. For instance, the relative improvement is the largest on the MovieLens dataset, indicating that in domains such as movie recommendation, users' long-term interests and preference characteristics have a more significant impact on recommendation performance.


\begin{table}[t]
\centering
\caption{Performance comparison of different profile generation methods (in accuracy).}
\label{tab:profile_perf}
\begin{tabular}{lccc}
\toprule
\textbf{Dataset} & \textbf{Without DPO} & \textbf{SFT} & \textbf{With DPO} \\
\midrule
MovieLens       & 50.9 & 51.1 & \textbf{53.0} \\
Yelp            & 64.9 & 66.2 & \textbf{70.4} \\
Amazon Books    & 59.1 & 63.7 & \textbf{70.4} \\
\bottomrule
\end{tabular}
\end{table}

\subsection{Ablation Study}

\begin{figure}[htbp]
    \centering
    \begin{minipage}[t]{0.45\textwidth}
        \centering
        \includegraphics[width=\textwidth]{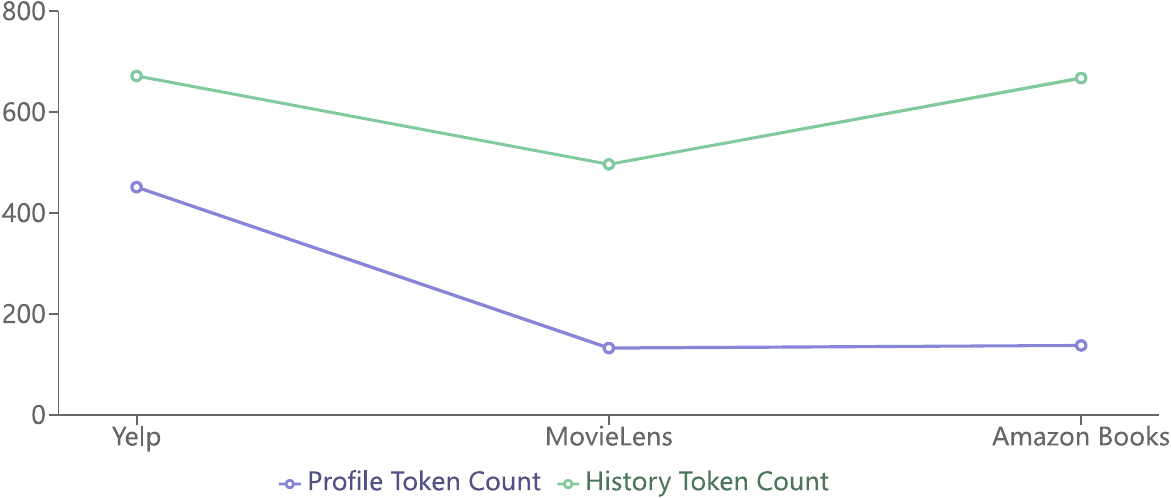}
        
    \end{minipage}
    
    \caption{Average token count comparison between profile and history on LLaMA3}
    \label{fig:tokens}
    \vspace{-3mm}
\end{figure}

\begin{figure*}[!t]
  \centering
  \includegraphics[width= 0.9\linewidth]{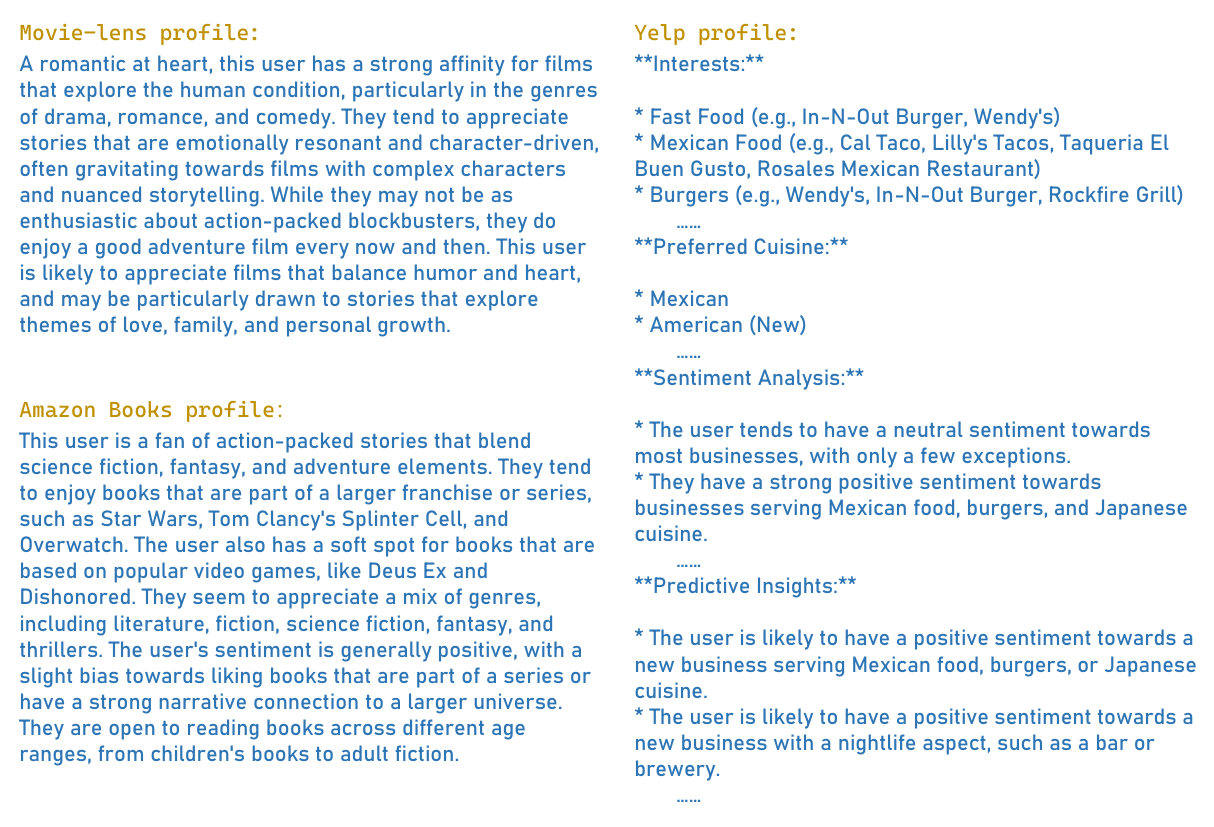}
  \caption{
   The open-format profiles generated by \textsc{LettinGo} vary across different datasets.
  }
  \label{fig:case}
\end{figure*}

\begin{figure}[htb]
  \centering
  \includegraphics[width=\linewidth]{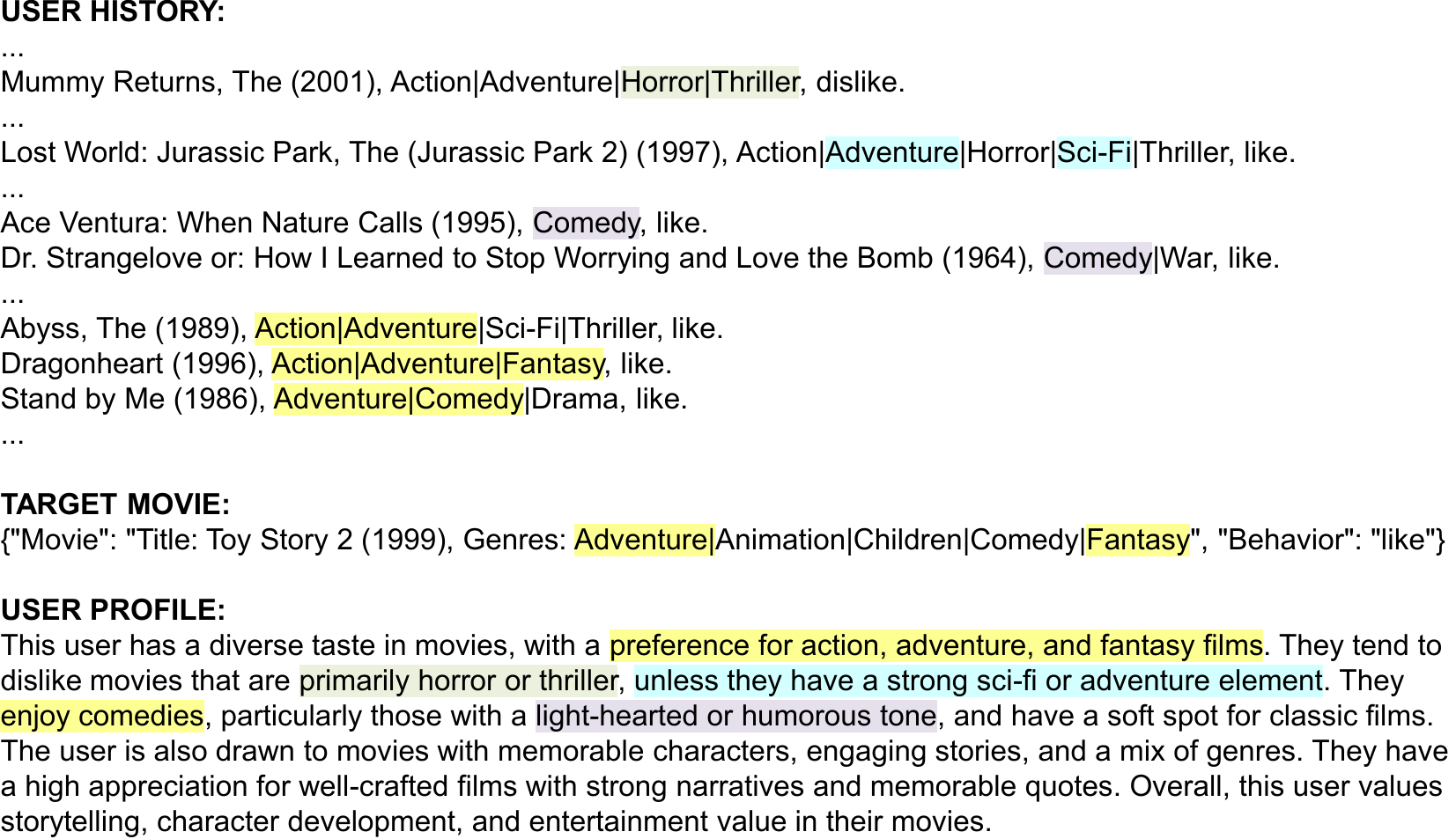}
  \caption{
    A sample where the prediction fails before using the profile but succeeds after using it. 
  }
  \label{fig:case2}
\end{figure}

\subsubsection{Effectiveness of optimization/alignment~(RQ2)} To investigate the impact of profile alignment on the final performance, we conducted ablation experiments on three datasets using the same experimental settings described in Section 4.1 on the LLama3 model. Specifically, we compared the performance before and after optimizing with DPO (Direct Preference Optimization). This analysis helps us understand how alignment contributes to generating more task-relevant and high-quality profiles.

As shown in Table~\ref{tab:profile_perf}, our DPO-based profile alignment method yields consistent and substantial improvements in accuracy across all three datasets. On the MovieLens-10M dataset, it improves accuracy by 2.1\% over the SFT baseline. The effect is even more pronounced on the Yelp dataset, with a 4.2\% improvement, and reaches its peak on the Amazon Books dataset, where accuracy increases by 6.7\%.

These results demonstrate that profile alignment using preference-based optimization effectively leverages feedback from downstream tasks to guide the profile generation model. By learning to produce profile formats that are better aligned with task-specific objectives, the method significantly enhances the quality of generated profiles, more accurately summarizes user preferences, and ultimately improves downstream recommendation performance.

\subsubsection{Impact of historical interaction length on profile quality~(RQ3)}We conducted another set of experiments to evaluate the impact of historical interaction length on the quality of the generated profiles. Specifically, we used different lengths of user interaction histories (e.g., the most recent 30, 50, and 70 interactions) to generate profiles, following the experimental settings described in Section 4.1. The results of these experiments are summarized in Table~\ref{table:main_result}.

Interestingly, the results did not align with our initial expectation that longer histories would always produce better profiles. Instead, the performance varied depending on the dataset and experimental conditions. This phenomenon can be attributed to the following factors:

\begin{itemize}[leftmargin=*]
    \item \textbf{Dataset Scale and Sparsity:} In datasets with sparse user interactions, shorter histories are often sufficient to capture the primary user preferences. In contrast, longer histories may introduce noise, reducing the quality of the profiles. On the other hand, in datasets with dense interactions, longer histories can provide richer semantic information, which benefits profile generation.

    \item \textbf{User Interest Dynamics:} User preferences often evolve over time. Shorter histories can better reflect recent preferences, which are typically more relevant for downstream tasks. However, longer histories may include outdated or irrelevant interactions, diluting the signal and potentially degrading the performance.

    \item \textbf{Model Capacity and Robustness to Noise:} The recommendation model's ability to process input data and handle noise also plays a critical role. For example, models with higher capacity can better extract meaningful signals from longer histories, while models with limited capacity may struggle with excessive input data, becoming more sensitive to noise.
\end{itemize}
These findings suggest that the choice of historical interaction length should be adapted dynamically, taking into account the characteristics of the dataset and the capacity of the recommendation model. This adaptive approach could help optimize the trade-off between capturing sufficient user preferences and avoiding noise, ultimately improving both profile quality and downstream recommendation performance.

\subsection{Case Study}

\subsubsection{Effectiveness of \ourmethod profile~(RQ4)} Figure~\ref{fig:tokens} presents the average token length of profiles generated by Llama3 using 30 interaction records from the test set, as well as the average token length of the same historical interaction records. The comparison demonstrates that, compared to directly feeding long historical records into the large language model, the generated profiles effectively reduce the context length. This reduction alleviates the challenges faced by the model in capturing information within long-context scenarios. As shown in Figure~\ref{fig:case2}, this is an example where the model makes incorrect predictions when using only 10 historical records, but successfully predicts when additional profile information is added. The parts with the same background color in USER HISTORY and USER PROFILE indicate the user preferences correctly summarized by the profile. For example, this user is very interested in movies of the action, adventure, and fantasy genres and also enjoys light and humorous comedies. However, they are not particularly fond of movies whose primary theme is horror or thriller, unless they contain strong elements of science fiction and adventure. In the prediction for the TARGET MOVIE, the profile also contributes useful information about the user's preference for science fiction, adventure, and comedy genres, which helps the model make a successful prediction.

In different domain-specific datasets, the user profiles generated by our method exhibit significant variations in format, reflecting the distinct demands of each domain and the diverse application scenarios of user profiling. As shown in Figure~\ref{fig:case}, the MovieLens user profile primarily adopts a narrative style, emphasizing emotional engagement and thematic preferences. This format is well-suited for analyzing subjective experiences and psychological inclinations. In contrast, the Yelp user profile employs a highly structured format, explicitly categorizing aspects such as interests, preferred cuisines, and sentiment analysis, making it more suitable for applications in machine learning and recommendation systems. Meanwhile, the Amazon Books user profile strikes a balance between narrative and structured formats, capturing users’ genre preferences and brand loyalty while maintaining a degree of readability and analytical flexibility. These differences in format highlight the need for domain-specific customization of user profiles, tailored to the unique characteristics of the dataset and its intended use cases. Our profiling approach, along with its corresponding optimizations, effectively addresses this requirement by providing a flexible framework that caters to both human interpretation and machine-driven analysis.

\subsection{Comparison with GPT-4o and Model Transferability}

To assess the competitiveness of our DPO-based profile generation method against closed-source models, we conducted a comparison with GPT-4o. On the MovieLens dataset, we used GPT-4o to generate user profiles and evaluated them with a LLaMA 3 8B Instruct predictor. Our method (DPO with 10H+70P) achieved 53.00\% accuracy and 51.69 F1, outperforming GPT-4o-generated profiles, which reached 52.80\% accuracy and 51.30 F1. These results indicate that our DPO-based generator produces more effective and consistent profiles in larger-profile settings.

We further evaluated the transferability of our approach by using Qwen2.5 7B Instruct for both profile generation and downstream prediction. As shown in Table~\ref{tab:qwen-transfer}, our method generalizes well across different model backbones, demonstrating consistent improvements as the profile size increases.

\begin{table}[H]
\centering
\caption{Performance of profile generation with Qwen2.5 7B Instruct on MovieLens.}
\label{tab:qwen-transfer}
\begin{tabular}{lcc}
\toprule
\textbf{Method} & \textbf{Accuracy} & \textbf{F1 Score} \\
\midrule
10H            & 52.50 & 50.23 \\
10H+30P        & \textbf{58.30} & \textbf{56.87} \\
10H+50P        & 56.64 & 55.16 \\
10H+70P        & 57.10 & 56.58 \\
\bottomrule
\end{tabular}
\end{table}

\section{Conclusion}

In this paper, we present \textsc{LettinGo}, a novel framework for flexible user profile generation that leverages large language models to generate flexible user profiles from long-term interaction histories, thereby enhancing recommendation system performance. Our framework consists of three key components: (1) Profile Exploration, which leverages both internal and external LLMs to generate diverse profile representations without format constraints; (2) Task-Driven Profile Evaluation, which assesses profile quality through downstream recommendation performance; and (3) Profile Preference Alignment, which optimizes profile generation through feedback-driven training using DPO. This design enables our framework to generate adaptive and high-quality profiles while maintaining flexibility in profile representation.Experimental results demonstrate that our method significantly outperforms existing baseline approaches across multiple datasets, validating the substantial value of flexible and adaptive user profiles in enhancing recommendation performance.

\newpage
\balance
\bibliographystyle{ACM-Reference-Format}
\bibliography{sample-base}
\appendix

\section{Baseline Prompts}\label{appendix:baselineP}

\textbf{KAR Prompt}

\begin{minipage}{\linewidth}
\begin{lstlisting}[breaklines=true, breakindent=0pt, xleftmargin=0pt]
Given the user's business reviewing history with sentiments over time, listed below: {user_history}, analyze the user's preferences, taking into account factors such as business name and categories. 
Provide clear explanations based on the details from the user's reviewing history and other pertinent factors.
\end{lstlisting}
\end{minipage}

\noindent\textbf{PALR Prompt}

\begin{minipage}{\linewidth}
\begin{lstlisting}[breaklines=true, breakindent=0pt, xleftmargin=0pt]
Your task is to use keywords to summarize user's preference based on history interations. The Output is an itemized list based on importance. The output template is {{1.KEY_WORD_1:"HISTORY_BUSINESS_1","HISTORY_BUSIN ESS_2"; 2.KEY_WORD_2:"HISTORY_BUSINESS_3"}}
The history businessed and their keywords and user' semtiment are:
{user_history}
\end{lstlisting}
\end{minipage}

\noindent\textbf{RLMRec Prompt}

\begin{minipage}{\linewidth}
\begin{lstlisting}[breaklines=true, breakindent=0pt, xleftmargin=0pt]
You will serve as an assistant to help me determine which types of businesses a specific user is likely to enjoy. I will provide you with information about businesses that the user has visited, as well as his or her sentiments of those businesses. Here are the instructions: 1. Each visited businesse will be described in the format with the following attributes: Title:the name of the business, Categories:the categories of the business, Sentiment:user semtiment toward business. 2. The information I will give you: INTERATION ITEMS: a list of JSON strings describing the items that the user has visited. Requirements: 1. Please provide your decision in JSON format, following this structure: {{ "summarization": "A summarization of what types of businesses this user is likely to enjoy" (if you are unable to summarize it, please set this value to "None") "reasoning": "briefly explain your reasoning for the summarization" }} 2. Please ensure that the "summarization" is no longer than 100 words. 3. The "reasoning" has no word limits. 4. Do not provided any other text outside the JSON string.
{user_history}
\end{lstlisting}
\end{minipage}

\end{document}